\documentclass[12pt]{article}                                     %
\usepackage{amsmath,amssymb,amsfonts,amsbsy}                          %
\usepackage{cite}                                                     %
\usepackage{graphicx}                                                 %
\usepackage{wrapfig}                                                  %
\usepackage{rotating}                                                 %
\usepackage{color}                                                    %
\usepackage[font=small,labelfont=bf,labelsep=period]{caption}         %
                                                                       
\RequirePackage{lineno} 
                                                                       
\usepackage{bbm} % blackboard bold                                    %
\usepackage{color}                                                    %
\usepackage{dsfont} % double stroke roman fonts                       %
\usepackage{latexsym} % a few additional special symbols              %
\usepackage{lscape} % allows single pages to be set landscape         %
\usepackage{mathrsfs} % alternative mathematical "script" font        %
\usepackage{morefloats} % solves excess floating figure problems      %
\usepackage{nonfloat} % for tables and figures that should not float  %
\usepackage{slashed} % slash notation for Dirac gamma-matrix products %

\begin{document}                                                       %
%%%%%%%%%%%%%
\graphicspath{{exper/vossen/}}
%%%%%%%%%%%%%
\addcontentsline{toc}{subsection}{{First Measurement of Interference 
Fragmentation Function in $e^+$ $e^-$ at Belle}\\
{\it B.B. Anselm Vossen}}

%%%%%%% please do not touch these! %%%%%%
\setcounter{section}{0}
\setcounter{subsection}{0}
\setcounter{equation}{0}
\setcounter{figure}{0}
\setcounter{footnote}{0}
\setcounter{table}{0}

\begin{center}
\textbf{FIRST MEASUREMENT OF THE INTEFERENCE FRAGMENTATION FUNCTION 
IN $e^+e^-$ AT BELLE}

\vspace{5mm}

\underline{A.~Vossen$^{\,1\,\dag}$}, R. Seidl$^{\,2}$, 
M.~Grosse Perdekamp$^{\,1}$,M. Leitgab$^{\,1}$, A. Ogawa$^{\,3}$ and
K.~Boyle$^{\,4}$ for the Belle Collaboration

\vspace{5mm}

\begin{small}
  (1) \emph{University of Illinois at Urbana Champaign} \\
  (2) \emph{RBRC (RIKEN BNL Research Center)} \\
  (3) \emph{BNL/RBRC} \\
  (4) \emph{RBRC}  \\
  $\dag$ \emph{E-mail: vossen@illinois.edu}
\end{small}
\end{center}

\vspace{0.0mm} % Don't laugh: it does change the spacing!
%\linenumbers
\begin{abstract}
A first measurement of the di-hadron interference fragmentation 
function of light quarks in pion pairs with the Belle detector is 
presented. The chiral odd nature of this fragmentation function 
allows the use as a quark polarimeter sensitive to the transverse 
polarization of the fragmenting quark. Therefore it can be used 
together with data taken at fixed target and collider experiments 
to extract the quark transversity distribution. A sample 
consisting of $711 \times 10^6$ di-hadron pairs was extracted 
from 661~$fb^{-1}$ of data recorded near the $\Upsilon(4S)$ 
resonance delivered by the KEKB $e^+$$e^{-}$ collider.
\end{abstract}

\vspace{7.2mm} 

\section{Introduction}

The di-hadron interference fragmentation function (IFF), 
suggested first by Collins, Heppelmann and Ladinsky~\cite{Collins:1993kq} describes the 
production of unpolarized hadron pairs in a jet from a 
transversely polarized quark. The transverse polarization is 
translated into an azimuthal modulation of the yields of hadron 
pairs around the jet axis. In addition the IFF is chiral odd and 
can therefore act as a partner for the likewise chiral odd quark 
transversity function. The resulting amplitude is chiral even and 
therefore leads to observable effects in semi deep inclusive 
scattering off a transversely polarized target \cite{hermesiff, 
compassiff} or in proton-proton collisions~\cite{yangiff} in 
which one beam is transversely polarized. Besides the 
fragmentation into two hadrons, the fragmentation of a 
transversely polarized quark into one unpolarized hadron can be 
used to extract transversity via the Collins effect. Measurement 
of this effect at Belle\cite{Collinsprd} made the first 
extraction of transversity possible\cite{alexei}. However as 
compared with the Collins fragmentation function, using the IFF 
to extract transversity exhibits a number of advantages. These 
are connected to the additional degree of freedom provided by the 
second hadron. It allows to define the azimuthal angle between 
the two hadrons as an observable in the transverse plane and at 
the same time integrate over transverse momenta of the quarks and 
hadrons involved. Because transverse momenta are integrated over 
collinear schemes in factorization and evolution can be used 
which are known and which do not need assumptions of the 
intrinsic transverse momenta\cite{Ceccopieri:2007ip}. Since the 
IFF is not a transverse momentum dependent function (TMD) it is 
universal and therefore directly applicable to SIDIS and proton 
proton data. From both types of experiment results are available 
\cite{hermesiff, compassiff,yangiff}. The results from SIDIS are 
indicating a non-zero IFF while the first analysis of the IFF 
effect at PHENIX~\cite{yangiff} opens a way to disentangle 
transverse spin effects in proton proton collisions. Because the 
IFF is not a TMD function, extraction of transversity times IFF 
is not dependent on a model of the transverse momentum 
dependence, which is the case in the measurement of the Collins 
effect~\cite{alexei}. Here transverse momentum in the final state 
originates from a convolution of quark distribution and 
fragmentation function. This leads also to the Sudakov 
suppression of the effect \cite{sudakov}. Technically, the 
extraction of the IFF from electron colliders is also easier, as 
the signal is not competing with asymmetries from QCD radiation. 
Also acceptance effects are smaller for the relative azimuthal 
angles of hadron pairs.

\section{Observables in $e^+e^-$ collisions}
\begin{figure}[h!t]
%% [number of text lines to wrap]{horizontal position: LRC}{figure width}
  \centering %% do not use \begin{center} ... \end{center}
  \vspace*{2mm} %% the vertical position may need tweaking
  \includegraphics[width=80mm]{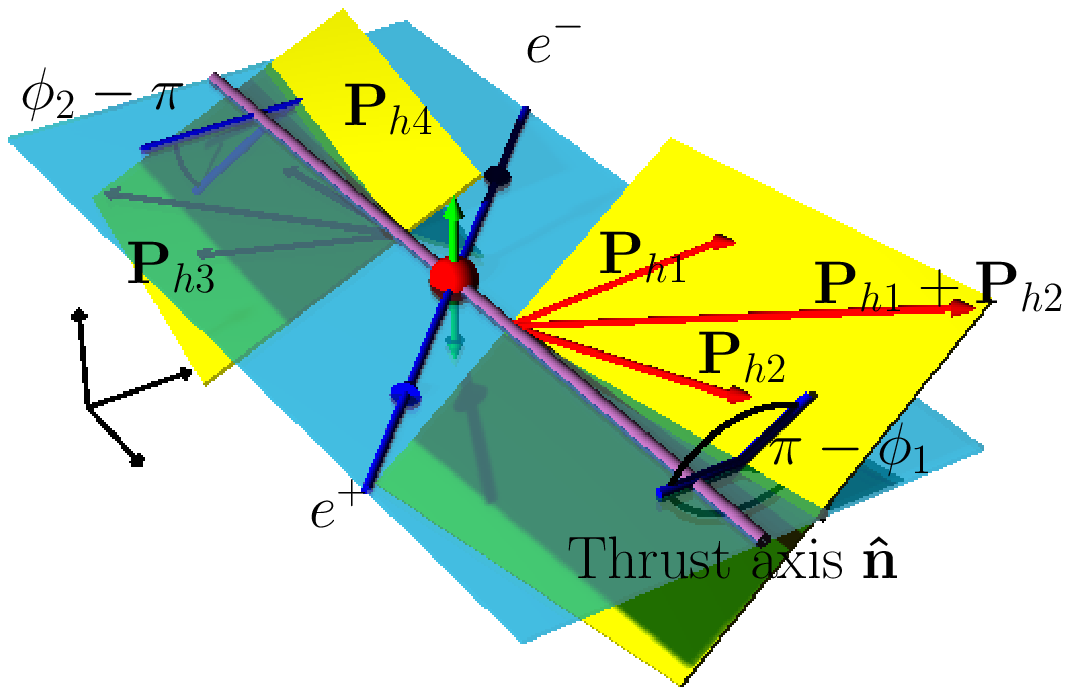}
  \caption{Azimuthal angle definition. Azimuthal angles $\phi_1$ and $\phi_2$ are defined relative to the thrust axis.}
  \label{fig:cooSys}
\end{figure}

The transverse polarization of the fragmenting quark leads to a 
cosine modulation of the azimuthal angle of the plane spanned by 
the two hadrons $h_1$, $h_2$ which is described by the vector 
$\mathbf{R}=\mathbf{P}_{h1}-\mathbf{P}_{h2}$ lying in the plane. 
Since the electron beams are unpolarized any effect that one 
would measure in one hemisphere of an event would average out. 
Instead one can make use of the fact that the spins of quark and 
anti quark in electron-positron annihilation are 100\% correlated. 
Thus the correlation of the azimuthal angles of the vectors 
$\mathbf{R}^\alpha$ in the hemispheres $\alpha\in\{1,2\}$ around 
the thrust axis with regard to the event plane is sensitive to 
the IFF. The measurement uses the center of mass system and 
defines the event plane as the plane which contains the beam axis 
$\hat{\mathbf{z}}$ and thrust axis $\hat{\mathbf{n}}$. Figure 
\ref{fig:cooSys} shows the coordinate system with the relevant 
quantities. With these the angles $\phi_\alpha$ between 
$\mathbf{R_\alpha}$ and the event plane can be expressed as
%fehlt noch referenz zur extraktion, e.g. boer

%%%%%%%% Fig. 1 %%%%%%%%

\begin{eqnarray}
\label{eq:phir}
\phi_{\{1,2\}}&=&\mathrm{sgn}\left[\hat{\mathbf{n}}\cdot (\hat{\mathbf{z}} \times \hat{\mathbf{n}} \times (\hat{\mathbf{n}}\times \mathbf{R}_{1,2})\}\right] \nonumber \\
&\times &\arccos \left(\frac{\hat{\mathbf{z}}\times \hat{\mathbf{n}}}{|\hat{\mathbf{z}}\times \hat{\mathbf{n}}|}
\cdot \frac{\hat{\mathbf{n}}\times \mathbf{R}_{1,2}}{|\hat{\mathbf{n}}\times \mathbf{R}_{1,2}|}
\right)\quad.
\end{eqnarray}
The product of the quark and anti quark interference 
fragmentation functions $H_1^\sphericalangle \cdot 
\bar{H}_1^\sphericalangle $ is then proportional to the amplitude 
$a_{12}$ of the modulation $\cos(\phi_1+\phi_2)$ of the di-hadron 
pair yields~\cite{Boer:2003ya}. The di-hadron fragmentation 
functions are dependent on the kinematic variables 
$m_\textrm{Inv}$, the invariant mass of the hadron pair, and 
$z=\frac{2E_h}{Q}$ the normalized energy of the hadron pair. Here 
$E_h$ is the energy of the hadron and $Q$ the absolute energy 
transferred by the virtual photon. Therefore the normalized 
azimuthal yield of di-hadron pairs can be described as
\begin{equation}
\label{eq:asy} N(y,z_1,z_2,m_1,m_2, \phi_1+ \phi_2)\propto a_{12}(y,z_1,z_2,m_1,m_2)\cdot \cos(\phi_1+\phi_2)
\end{equation}
with
\begin{equation}
a_{12}(y,z_1,z_2,m_1,m_2) \sim B(y)\cdot 
\frac{\sum_q e_q^2 H_1^\sphericalangle(z_1,m_1) 
\bar{H}_1^\sphericalangle(z_2,m_2)} {\sum_q e_q^2 
D_1(z_1,m_1)\bar{D}_1(z_2,m_2)}.
\end{equation}
Here the sum goes over all quark flavors $q$ and 
$B(y)=\frac{y(1-y)}{\frac{1}{2}-y+y^2}\stackrel{\textrm{CM}}{=} 
\frac{\sin^2\theta}{1+\cos^2\theta}$ is the kinematic factor 
describing the transverse polarization of the quark-anti quark 
with regard to its momentum. The polar angle $\theta$ is defined 
between the electron axis and the thrust axis as shown in 
fig.~\ref{fig:cooSys}. The yields are dependent on $z_1$, $z_2$, 
$m_1$, $m_2$, the fractional energies and invariant masses of the 
hadron pairs in the first and second hemisphere, and $y$. 
In the following the yields are integrated over $y$ in the limits of the fiducial cuts on $\theta$
introduced in section \ref{sec:dataSelection}.
The labeling of the hemisphere is at random, and it is experimentally not 
possible to distinguish between quark and anti quark 
fragmentation.
Further information about the IFF can be learned from the 
dependence on the decay angle $\theta_h$ in the CMS of the two 
hadrons produced. Using a partial wave decomposition to isolate 
components that contain the interference between waves with one 
unit difference in angular momentum, one expects a dependence on 
$\sin\theta_h$ which gives the interference term between s and p 
waves. This term is expected to dominate at Belle kinematics and 
is favored by the acceptance. Since the acceptance is symmetric 
around $\theta_h=\frac{\pi}{2}$ the p-p contribution proportional 
to $\cos\theta_h$ should average out. Table~\ref{tab:avgKin} 
shows that this is approximately the case.

\subsection{Models}
As described in the previous section the IFF is an interference 
effect between hadrons, here pions, created in partial waves with 
a relative angular momentum difference of one. The dominant 
contribution being from the s-p interference 
term~\cite{accessingTransvWIFF, twoHadIFFModCalc}. Therefore 
information about the IFF can be gained from a partial wave 
analysis for di-pion production \cite{phaseShiftModel}. Here the 
data from \cite{phaseShiftData} suggest a phase shift around the 
$\rho$ mass, leading to a sign change in the fragmentation 
function. In some models\cite{modDiHadFF} the location of the 
phase shift around the mass of the $\rho$ meson is caused by the 
interference of pion pairs produced in a p wave coming from the 
decay of the spin one $\rho$ which interferes with the 
non-resonant background. Based on this model and estimation of 
particle yields from simulations, Bacchetta, Ceccopieri, 
Mukherjee and Radici \cite{Radici} made model predictions for the 
magnitude of IFF Asymmetries at Belle.
%These are shown in fig. \ref{fig:baccPred}.
Again, a strong dependence on the invariant mass is predicted, 
with a maximum around the rho mass. The asymmetries are expected 
to rise with $z$ due to the preservation of spin information 
early in the fragmentation.

\subsection{The Belle experiment}
The Belle detector, located at the KEKB asymmetric energy 
$e^+e^-$ collider is described in detail in \cite{Belle}. For the 
purpose of this study it is important that a high number of 
events is recorded, a good particle identification allows to 
identify pions up to high values of $z$ and that the detector is 
hermetic to minimize acceptance effects. Both is well fulfilled 
by the Belle detector. It is located at the collision point of 
the 3.5 GeV $e^+$ and 8 GeV $e^-$ beam and is almost symmetric in 
the center of mass system of the beams. It is a large solid angle 
magnetic spectrometer consisting of barrel and end caps parts. 
For a more homogeneous acceptance function only the barrel part 
was used for this analysis. It comprises a silicon vertex 
detector, a 50-layer drift chamber and an electromagnetic 
calorimeter (CsTI) in a magnetic field of 1.5T. Particle 
identification over a wide range of momenta is done using an 
array of aerogel Cherenkov counters, a barrel-like arrangement of 
time-of-flight scintillation counters and an instrumented iron 
flux return yoke outside of the coil to detect $K_L^0$ mesons and 
muons.

\subsection{Data selection and Asymmetry extraction}
\label{sec:dataSelection}
 From the 661 $\textrm{fb}^{-1}$ data sample roughly 589 
$\textrm{fb}^{-1}$ were taken on the $\Upsilon(4S)$ resonance and 
73 $fb^{-1}$ taken in the continuum 60 MeV below. Because the 
thrust cut used to select events with a two jet topology 
containing light and charm quarks also rejects events in which B 
mesons were produced, on-resonance and continuum data can be 
combined. The thrust is defined as $T=\frac{\sum_i |p_i \cdot 
\hat{n} |}{\sum_i|p_i|}$ and $\hat{\mathbf{n}}$ is direction of 
the thrust chosen such that T is maximal. B meson events produced 
on the $\Upsilon$ resonance have a spherical shape, since their 
high mass does not allow for large kinetic energy. On the other 
side, light quark anti quark pairs have a more two jet like 
topology.. An applied thrust cut of $T>0.8$ reduces the 
contamination with B events to an order of 2\% \cite{Collinsprd}. 
The inversion of the thrust cut selects events that don't have a 
clear two jet topology and that are contaminated by $\Upsilon$ 
decays into B mesons. This leads to a decrease of the asymmetry. 

As described earlier, only the barrel region of the detector is 
used in the analysis. Therefore the thrust axis is required to 
lie in a region well contained within it. This translates into a 
cut on the z component of the thrust axis $|\mathbf{n}_z|< 0.75$, 
which also allows for the particles of the jet around the axis to 
be reconstructed in the barrel. Further cuts on the event level 
are a reconstructed energy of at least 7 GeV to reject $e^+e^- 
\rightarrow \tau^+\tau^-$ and to reliably reconstruct the thrust 
axis. The later is computed using all charged tracks and photons 
passing some minimum energy cuts. A mean deviation of 135~mrad 
with a RMS of 90~mrad of the thrust axis from the real quark-anti 
quark axis is computed from simulations.

Only events were selected which satisfy a vertex cut of 2~cm in 
the radial and 4~cm in the beam direction. On the track level the 
fiducial cuts to reduce acceptance effects are a constraint to 
the barrel region of the detector using a cut on the polar angle 
$\theta$ in the laboratory system of $-0.6 <\cos(\theta)<0.9$ 
which translates to an almost symmetric cut in the CMS. To make 
sure that the azimuthal range of tracks around the thrust axis is 
not biased, only tracks are chosen that have at least 80\% of 
their energy along the thrust axis. This restricts tracks to be 
within a cone that is entirely contained in the acceptance and 
reduces false asymmetries considerably as shown later in 
sec.~\ref{sec:sysSt}. No false asymmetries from this cut is 
expected and since most of the energy of the jet is contained 
within the jet no significant dilution of the asymmetries either. 
Only tracks above a minimal fractional momentum $z>0.1$ are 
considered. They have to be positively identified as pions. These 
tracks are then sorted into two hemispheres according to the sign 
of their momentum projection on the thrust axis. All possible 
pairs of $\pi^+ \pi^-$ in the same hemisphere are selected and 
the angle $\phi_\alpha, \alpha\in\{1,2\}$ computed. To this end 
the vector $\mathbf{R}_\alpha=\mathbf{P}_1-\mathbf{P}_2$ for each 
pion pair in hemisphere $\alpha$ is formed and the azimuthal 
angle around the thrust axis with respect to the event plane 
computed according to eq. \ref{eq:phir}. Charge ordering of 
$h_1$, $h_2$ in the computation of $\mathbf{R}$ is always the 
same so that the effect is not averaged out. A weighting of the 
hadron momentum vector with the inverse fractional energy $z$ as 
suggested by \cite{Artruiff} only led to differences in the 
thrust axis within numerical uncertainties. For the computation 
of the yields in a specific $\phi_1+\phi_2$ bin all combinations 
of pairs in the two hemispheres are considered. Due to the cone 
cut described earlier, the number of hadrons with a false 
hemisphere assignment is negligible. From Monte Carlo studies a 
signal purity for di-pion pairs (4 particles) of better than 90\% 
over the whole kinematic range is obtained.

\subsection{Systematics studies}
\label{sec:sysSt} In order to determine the systematic error on 
the measurement, simulation and real data was used to determine 
the contribution of detector effects and competing physical 
processes to false asymmetries or a dilution of the measurement. 
The studies that lead to the biggest contribution to the 
systematic error are the study of false asymmetries in simulation 
and real data in which no asymmetry is expected due to a wrong 
assignment of the thrust axis or the hemispheres of the particle 
pairs. Any false asymmetries, together with their statistical 
errors were added to the systematic error.

Since the IFF effect is not included in the Pythia event 
generator used, checking for the asymmetries in fully simulated 
data allows to estimate effects of the detector acceptance and 
efficiencies on the asymmetry. Table \ref{tab:mcRes} shows the 
false asymmetries extracted from a full simulation of the 
detector using GEANT.

Another source for false asymmetries are events reconstructed 
from real data in which the asymmetries are averaged out. This is 
the case for mixed events in which the angles $\phi_1$ and 
$\phi_2$ are taken from different events or events in which the 
angels are computed for pion pairs in the same hemisphere. The 
later case leads to asymmetries due to phase space restrictions, 
which could be reproduced, the former to asymmetries in the order 
of one per-mille, shown in table~\ref{tab:mixing}, which has been 
added to the systematic error. Contributions from higher 
harmonics in the fits to the cosine modulation are under one 
per-mille.

Due to the smearing of the thrust axis reconstruction with 
respect to the true quark-anti quark axis, the extracted 
asymmetries are diluted. Since this dilution could be reproduced 
in weighted Monte Carlo using the observed smearing of the thrust 
axis, it was corrected for.

More physics motivated systematic checks concern the dependence 
on the various kinematic factors. The dependence on the kinematic 
factor  $\frac{\sin^2 \theta}{1+\cos^2\theta}$ should be linear, 
as should be the dependence on $\sin\theta_h$, if the effect is 
dominated by the s-p interference term. Both could be validated.

Even though the Belle detector is very stable over time, checks 
have been done to determine the compatibility of data taking 
periods and data taken on and off the $\Upsilon$ resonance. To 
this end, the $\chi^2$ of each fit has been computed and all 
values have been fitted by an appropriate $\chi^2$ distribution. 
The result of these tests show very good compatibility.

We checked for correlations in the data that might lead to false 
error estimates by breaking up weighted Monte Carlo data in 500 
small chunks and comparing the mean error of the extracted 
asymmetries to their variation. The results are compatible, so no 
systematic error was assigned.

\begin{figure}[h!t]
  \begin{center}
  \begin{tabular}{cc}
  \includegraphics[width=0.47\textwidth]{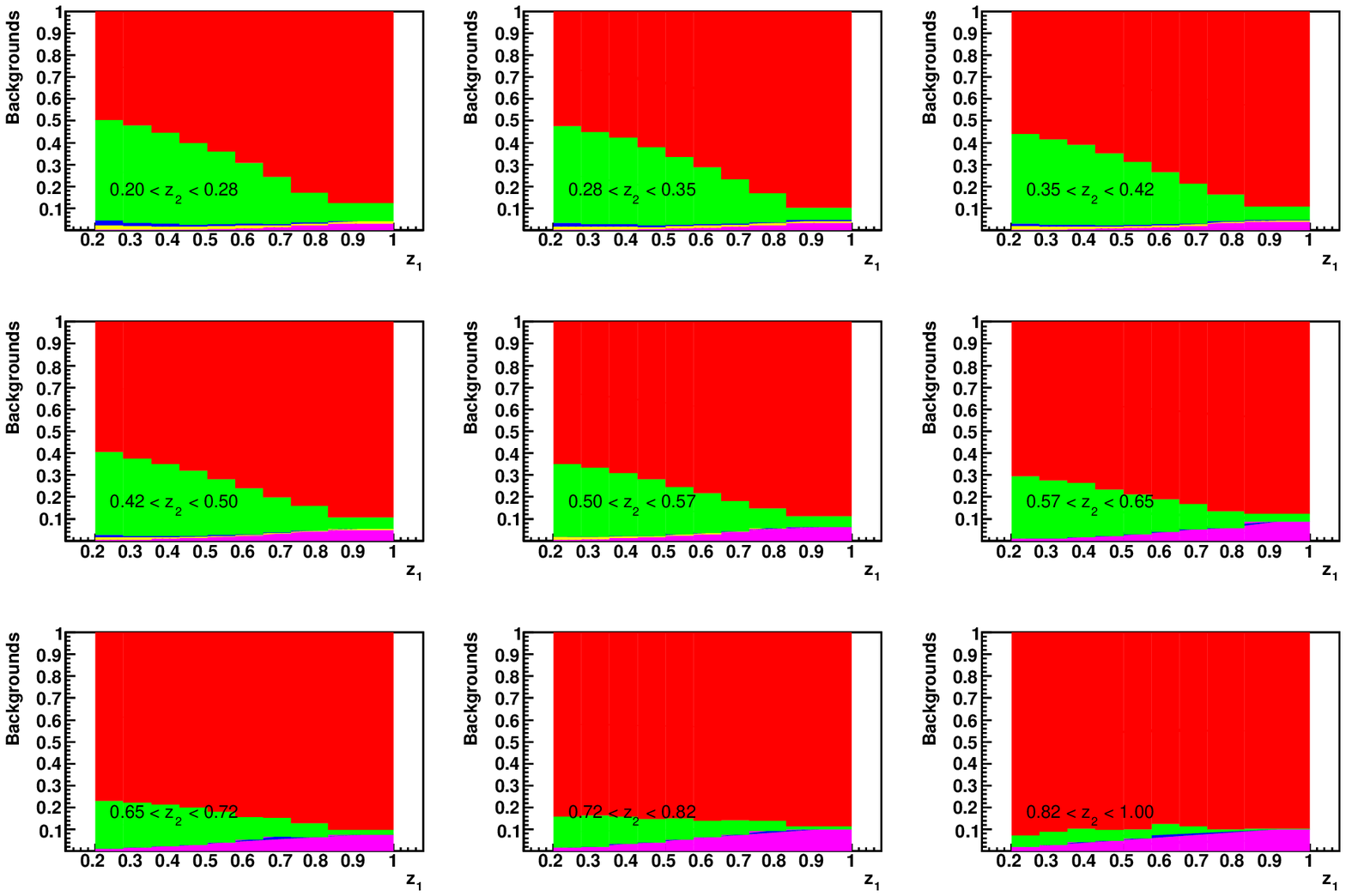} & \includegraphics[width=0.47\textwidth]{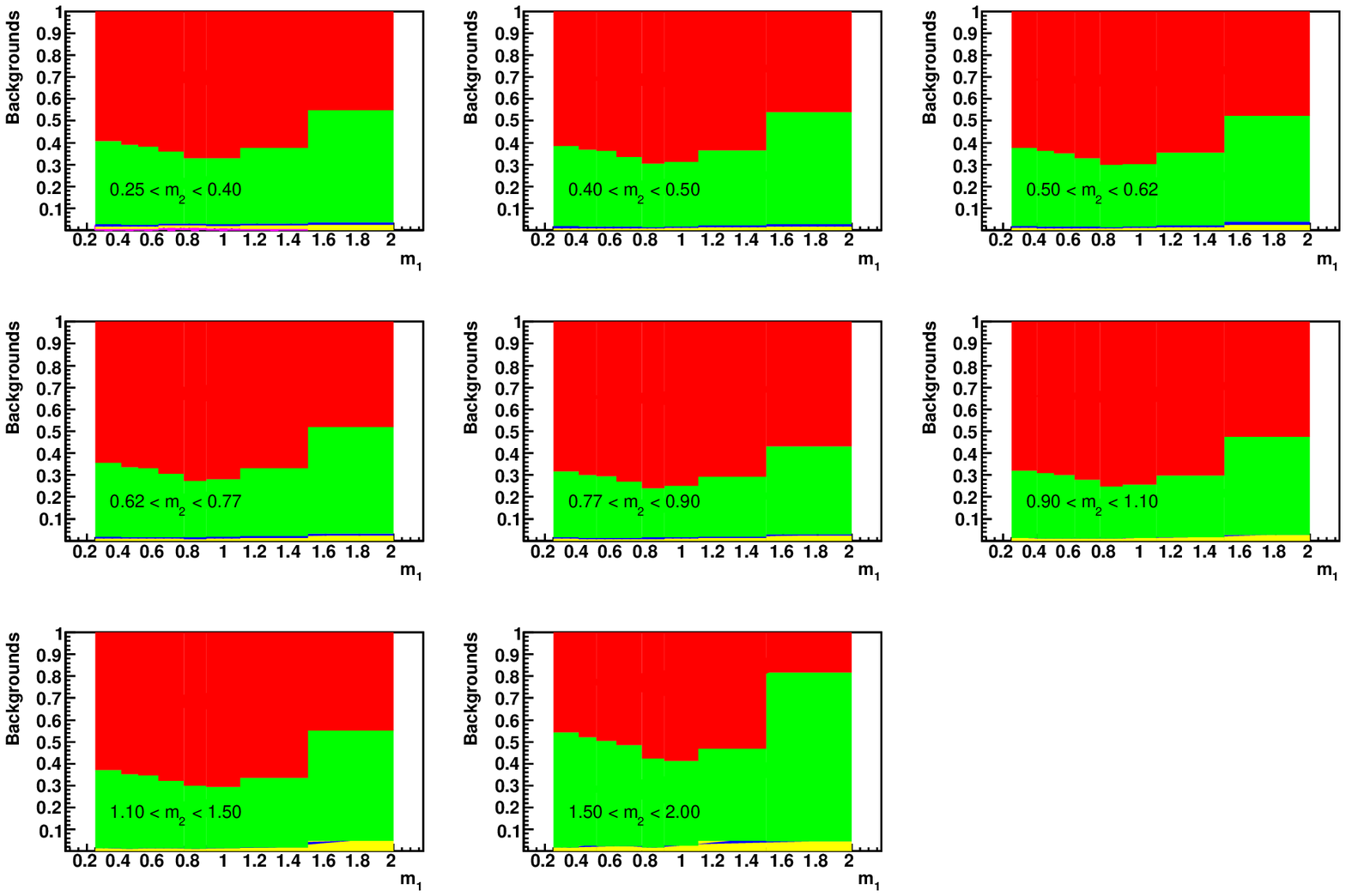}\\
  \textbf{(a)} & \textbf{(b)} 
  \end{tabular}
  \caption{\footnotesize Relative process contributions from light quark-anti quark events (red), charm events (green), charged B meson pairs (blue), neutral B meson pairs (yellow) and $\tau$ pairs (purple) as a function of $z_2$ for all $z_1$ bins (a) and as a function of $m_2$ for all $m_1$ bins (b)}
  \label{fig:processContributions}
  \end{center}
\end{figure}

For the interpretation of the results the respective fraction of 
processes contributing to the asymmetries are very important. 
These have been estimated from Monte Carlo simulations and are 
shown in fig.~\ref{fig:processContributions}. It is evident that 
the contribution from $\Upsilon$ decays has been almost 
eliminated by the thrust cut. There are also marginal 
contributions from $\tau$ pairs. For these the false asymmetries 
are compatible with zero and they have been added with their 
statistical error to the overall systematic error. Besides the 
contribution from light quark pairs there is a considerable 
contribution from charm quarks. This contributions decreases with 
increasing $z$ which can be understood, since charmed mesons have 
to undergo an additional decay before they can contribute to the 
pion asymmetries. This decay can also cause the invariant mass 
dependence namely a general decrease with higher masses. However, 
the highest bin shows a very high charm contribution, in some 
bins more than half of the events, which is not yet understood. 
The asymmetry results for these bins indicate that the IFF for 
charm quarks is non-vanishing and of similar magnitude as that of 
lighter quarks.

\subsection{Results}
\begin{figure}[h!]
%% [number of text lines to wrap]{horizontal position: LRC}{figure width}
  %\centering %% do not use \begin{center} ... \end{center}
  \begin{center}
  \begin{tabular}{c}
  \vspace*{-8mm} %% the vertical position may need tweaking (width used to be 80mm
  \includegraphics[width=0.85\textwidth]{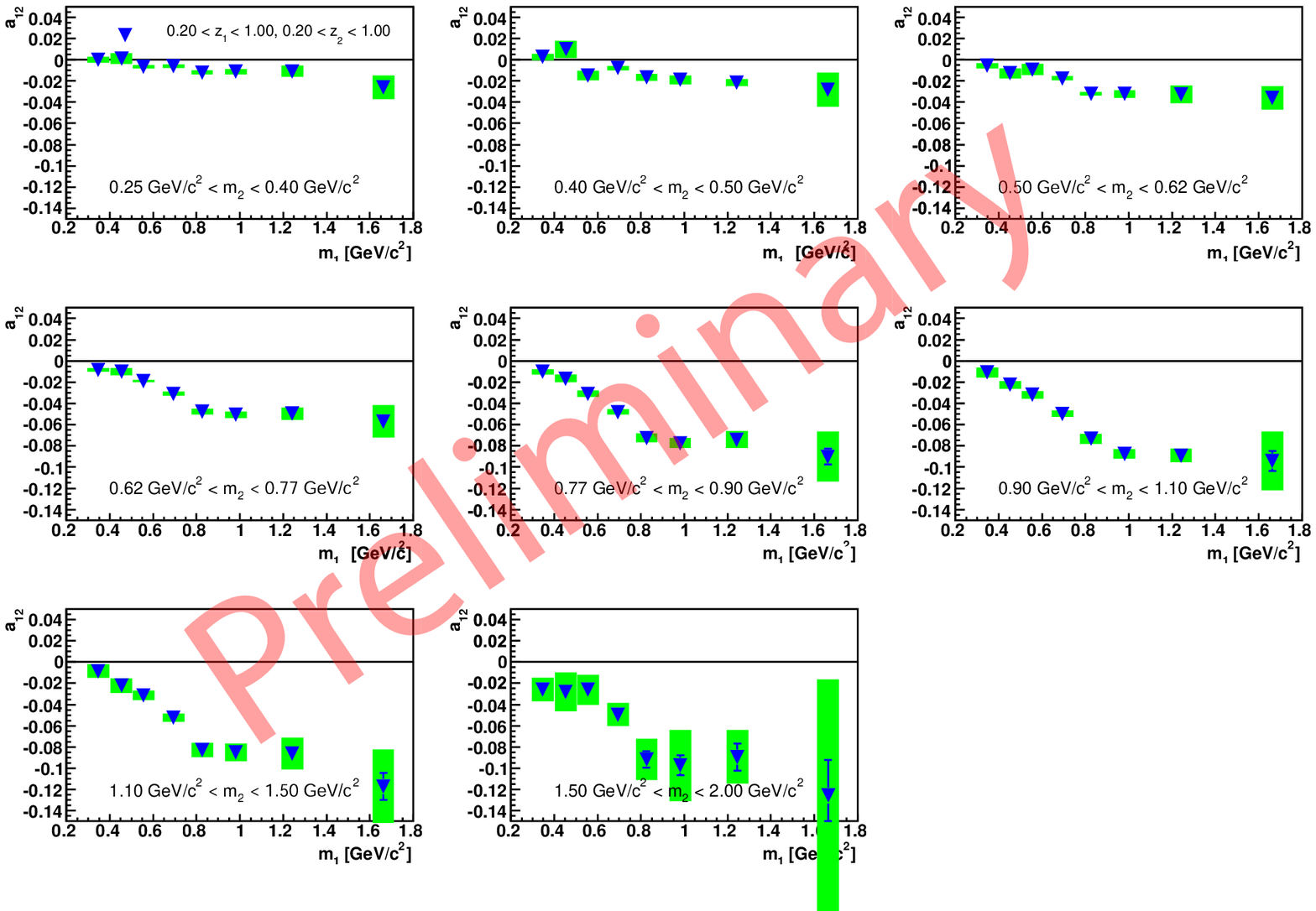} \\
  %\vspace{-65mm}\\
  %\rotatebox{45}{\textcolor{red}{{\Huge Preliminary}}}
  %\vspace{50mm}\\
  %\vspace{5mm}\\
  \textbf{ (a) }\\
  \includegraphics[width=0.85\textwidth]{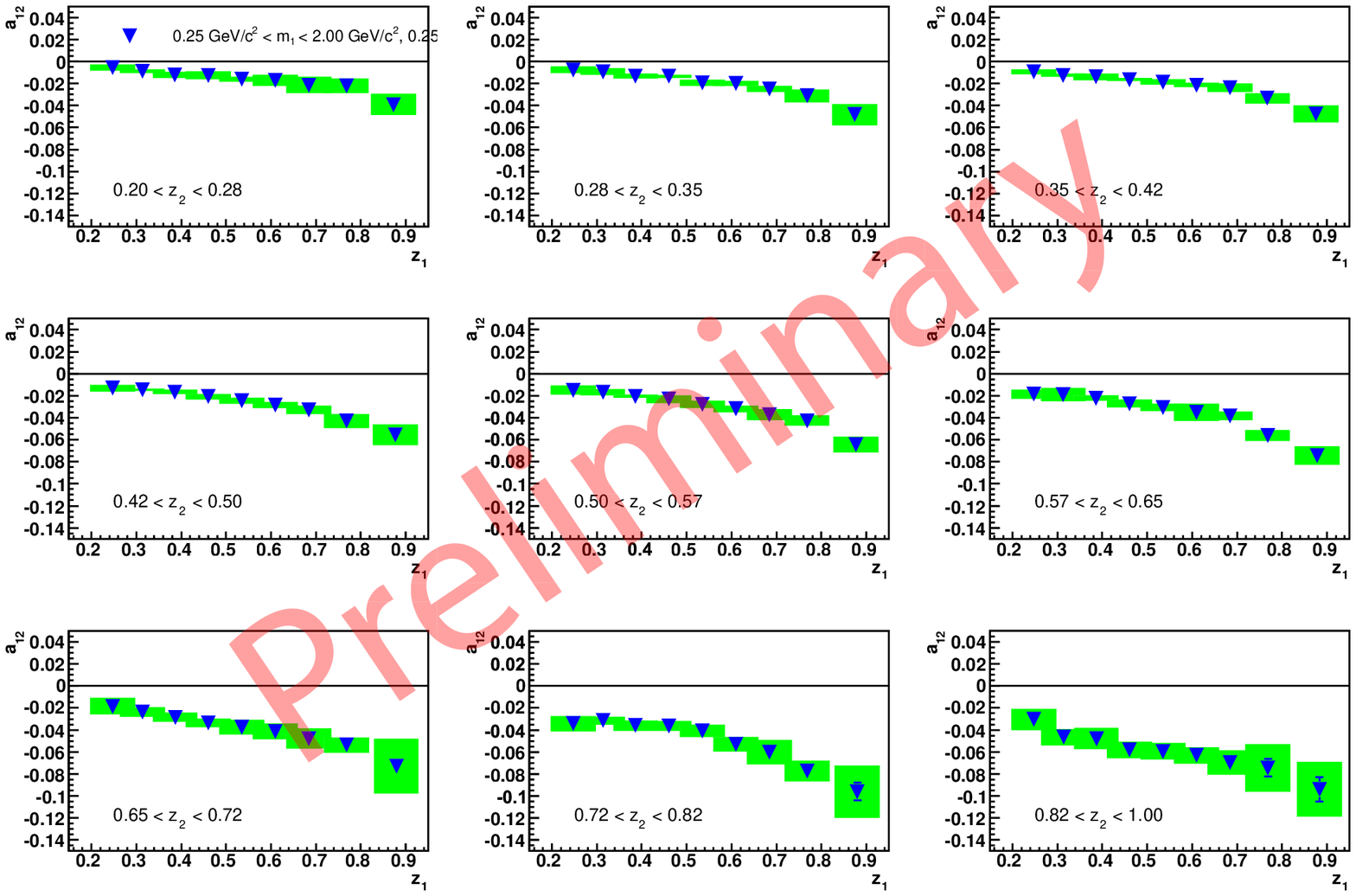}\\
  %\vspace{-70mm}\\
  %\rotatebox{45}{\textcolor{red}{{\Huge Preliminary}}}
  %\vspace{75mm}\\
  %\vspace{5mm}\\
   \textbf{(b)} 
  \end{tabular}
  \caption{\footnotesize Results for the $a_{12}$ modulations in a 
symmetric $8 \times 8$ binning in $m_1$, $m_2$ for $m_{\{1,2\}}$ 
between 0.25~GeV and 2~GeV (a) and for a symmetric $9 \times 9$ 
binning in $z_1$, $z_2$ for $z_{\{1,2\}}$ between 0.2 and 1 (b). 
The statistical error is shown in blue, the systematic error in 
green}
  \label{fig:resMZ}
  \end{center}
\end{figure}

The results obtained for the $a_{12}$ asymmetry as defined in 
eq.~\ref{eq:asy} are shown in figs.~\ref{fig:resMZ} binned in the 
invariant masses $m_1$, $m_2$ of the hadrons pairs in the first 
and second hemisphere and their fractional energies $z_1$, $z_2$, 
respectively. Table~\ref{tab:avgKin} shows the integrated 
asymmetries and the averaged kinematic observables. The extracted 
asymmetries are large, especially when considering that a product 
of the IFF for quark and anti quarks is measured. As expected the 
magnitude of the effect rises with $z$. However, the invariant 
mass behavior does not match model predictions from \cite{Radici}.
% which are shown in fig.~\ref{fig:baccPred}.
But these model predictions are only available in leading order 
and heavily dependent on simulations which were tuned for the 
SIDIS experiment HERMES at a center of mass energy of roughly 
7~GeV. The asymmetries rise up to around the mass of the $\rho$ 
but then plateau instead of decreasing again. A sign change of 
the IFF can therefore not be confirmed. However bins with high 
invariant masses receive also considerable contributions from 
charm quarks as shown before.

\subsection{Summary and Outlook}
The first direct measurement of the interference fragmentation 
function using 661~$fb^{-1}$ of data recorded at the Belle 
experiment has been presented. The asymmetries are large, up to 
10~\%, which would correspond to an IFF contribution of over 30\%. 
Models predicting a sign change or a decrease for invariant 
masses higher than the $\rho$ meson's could not be confirmed. 
Charm quarks play a significant role at high invariant masses 
They seem to introduce an IFF asymmetry of similar magnitude as 
light quarks and further studies are under way to determine the 
charm quark contribution to the asymmetries. The analysis 
presented here should enable a combined analysis to extract the 
transversity distribution of data taken in SIDIS, proton-proton 
and $e^+ e^-$. This is very desirable due to the various 
advantages as compared with the extraction via the Collins effect 
and its complementarity. In proton-proton collisions the use of 
the IFF effect to access transversity is especially helpful, 
since it can help disentangle different contributions to the 
measured transverse single spin asymmetries $A_N$. Our plans for 
the future contain also an extraction of other particle 
combinations, namely those including neutral pions and charged 
kaons. Furthermore an extraction of the unpolarized yields is 
planned to facilitate the extraction of the IFF.

%\begin{figure}[h!t]
%% [number of text lines to wrap]{horizontal position: LRC}{figure width}
  %\centering %% do not use \begin{center} ... \end{center}
 % \begin{center}
 % \begin{tabular}{cc}
 % \vspace*{-8mm} %% the vertical position may need tweaking
 % \includegraphics[width=80mm]{images/final_asypanel_val0_mz_bin5.eps} & %\includegraphics[width=80mm]{images/final_asypanel_val0_z_bin4.eps} \\
%  \vspace{5mm}\\  
%  \textbf{(a)} & \textbf{ (b) } 
%  \end{tabular}
%  \caption{Prefactor adjusted $a_{12}$ modulations for a $5 \times 10$ $z_1$, $m_1$ binning as a function of $m_1$ for the $z_1$ bins (a) and for a $5 \times 10 $ $z_1$, $z_2$ binning  as a function of $z_1$
%  for the $z_2$ bins (b) as compared to the theory predictions from \cite{Radici}. Only the 4 $z_1$ bins where predictions and data exist are shown.}
%  \label{fig:baccPred}
%  \end{center}
%\end{figure}

\begin{table}[th]
\begin{center}
\caption{\footnotesize MC results averaged over all $z$ bins in \%.
\label{tab:mcRes}}
\begin{tabular}{cc rr }\hline
sample &species &\multicolumn{2}{c}{$z_1,z_2$-Asymmetries}\\
%& &\multicolumn{2}{c}{Method 1}&\multicolumn{2}{c|}{Method 2}\\
& &$\langle a_{12}\rangle$&$\langle a_{12R}\rangle$\\ \hline
\multicolumn{4}{c}{No opening cut}\\ \hline
uds $4\pi$& $\pi\pi$&$-0.089\pm0.008$ &$-0.108\pm0.008$ \\
uds acceptance & $\pi\pi$&$-0.488\pm0.011$ &$-0.490\pm0.011$\\
uds MC rec. & $\pi\pi$ & $-0.394\pm0.013$ &$-0.418\pm0.013$ \\
charm rec. & $\pi\pi$ & $-0.446\pm0.041$ &$-0.388\pm0.044$ \\ \hline
\multicolumn{4}{c}{With opening cut of 0.8}\\ \hline
uds $4\pi$& $\pi\pi$&$-0.038\pm0.013$ &$-0.035\pm0.013$ \\
uds acceptance & $\pi\pi$&$-0.112\pm0.016$ &$-0.113\pm0.016$ \\
uds MC rec. & $\pi\pi$ & $0.012\pm0.019$ &$0.008\pm0.019$ \\
charm rec. & $\pi\pi$ & $0.006\pm0.040$ &$0.027\pm0.040$ \\
\hline
\end{tabular}
\end{center}
\end{table}

\begin{table}[ht]
\begin{center}
\caption{\footnotesize Mixing results averaged over the $z$ 
binning in \%. The results integrated over other binnings are 
nearly identical.\label{tab:mixing}}
\begin{tabular}{c rr }\hline
sample &\multicolumn{2}{c}{$z_1,z_2$-Asymmetries}\\
%& &\multicolumn{2}{c}{Method 1}\\\hline
& $\langle a_{12}\rangle$&$\langle a_{12R}\rangle$\\
uds $4\pi$& $0.070\pm0.013$ & $0.030\pm0.013$ \\
uds acceptance & $0.020\pm0.016$ &$-0.021\pm0.016$ \\
uds rec. &  $0.091\pm0.019$ &$0.087\pm0.019$ \\
charm rec. & $-0.024\pm0.040$ &$-0.017\pm0.040$ \\
Data &  $-0.019\pm0.017$ &$-0.012\pm0.017$ \\ \hline
\end{tabular}
\end{center}
\end{table}

\begin{table}[h!t]
\begin{center}
\caption{\footnotesize Integrated asymmetries and average kinematics.
\label{tab:avgKin}}
\begin{tabular}{c r}\\\hline
$\langle z_1 \rangle,\langle z_2 \rangle$ &0.4313\\
%$\langle z_2 \rangle$ &0.4312\\   
$\langle m_1 \rangle,\langle m_2 \rangle$ &0.6186\\   
%$\langle m_2 \rangle$ &0.6186\\   
$\langle \sin^2\theta/(1+\cos^2\theta) \rangle$ &0.7636\\   
$\langle \sin\theta_{h} \rangle$ & 0.9246\\   
%$\langle \sin\theta_{d2} \rangle$ &0.9246\\   
$\langle \cos\theta_{h} \rangle$ &0.0013\\   
%$\langle \cos\theta_{d2} \rangle$ &0.0014\\   
$a_{12}$&$ -0.0199 \pm  0.0002 \pm  0.0009$\\ \hline 
%$a_{12}$&$ -0.0199248 \pm  0.000188782 \pm  0.000879192$\\
%$a_{12R}$&$   -0.0177618 \pm  0.000175529 \pm  0.0007626$\\ \hline 
\end{tabular}
\end{center}
\end{table}

%%%%%%%%%%%%%%%%%%%%%%%%
%\clearpage

\end{document}